\documentclass[reprint,aps,prl,twocolumn,groupedaddress]{revtex4-1}
\usepackage{graphics}
\usepackage{graphicx}
\usepackage{epsfig}
\usepackage{amssymb}
\usepackage{amsmath}
\usepackage{epstopdf}
\DeclareGraphicsExtensions{.pdf,.eps,.png,.jpg,.mps}

\begin{document}
\title{Bound states in the continuum in graphene quantum dot structures}
\author{J. W. Gonz\'alez$^{1}$, M. Pacheco$^{1}$\cite{email}, L. Rosales$^{2}$
and P. A. Orellana$^{3}$}
\address{$^{1}$Departamento de F\'{i}sica, Universidad  T\'{e}cnica Federico Santa
Mar\'{\i}a, Casilla 110 V, Valpara\'{i}so, Chile}
\address{$^{2}$Instituto de F\'{i}sica, Pontificia Universidad Cat\'{o}lica
de Valpara\'{\i}so, Casilla 4059, Valpara\'{i}so, Chile}
\address{$^{3}$Departamento de F\'{i}sica, Universidad Cat\'{o}lica del Norte,
Casilla 1280, Antofagasta, Chile}

\date{\today}

\begin{abstract}
The existence of bound states in the continuum was predicted at the
dawn of quantum mechanics by von Neumann and Wigner. In this work we
discuss the mechanism of formation of these exotic states and the
feasibility to observe them experimentally in symmetrical
heterostructures composed by segments of graphene ribbons with
different widths forming a graphene quantum dot. We identify the
existence of bound states in the continuum in these graphene quantum
dot systems by means of local density of
states and electronic conductance calculations.
\end{abstract}

\maketitle

The new material denominated graphene is a single layer of carbon
atoms which can be fabricated by different methods like mechanical
peeling or epitaxial growth \cite{Novoselov,berger1,berger2}.
Nanoribbons are stripes of graphene  which can be obtained
through high-resolution lithography \cite{chinos}, by controlled
cutting processes \cite{Ci} or by unzipping multiwalled carbon
nanotubes \cite{Kosynkin}.
The electronic
behavior of all these nanostructures is mainly determined by their
geometric confinement which allows the observation of quantum
effects such as quantum interference effects, resonant tunneling
and localization effects. The possibility to control these quantum
effects, by applying external perturbations to the nanostructures or
by modifying the geometrical confinement, could be used to develop
new technological applications, such as graphene-based composite
materials \cite{stankovic},  molecular sensor devices
\cite{schedin,Rosales} and  nanotransistors \cite{Stampfer}.

An interesting feature exhibited by certain confined nanostructures, such as quantum dots systems,
is the presence of bound states in the continuum (BICs).  Their existence was predicted at the dawn of
quantum mechanics by von Neumann and Wigner \cite{boundstate1} for certain spatially
oscillating attractive potentials, for a one-particle Schr\"{o}dinger
equation. Much later, Stillinger and Herrick \cite{stillinger}
generalized von Neumann's work by analyzing a two-electron problem,
they found BICs were formed despite the interaction between
electrons. The occurrence of BICs was discussed in
a system of coupled Coulombic channels and, in particular, in a
Hydrogen atom in a uniform magnetic field \cite{friedrich}. BICs have also shown to be present in the electronic
transport in mesoscopic structures \cite{schult,zhen-li,nockel,olendski, rotter,loreto}. More
recently, exploiting the analogy between electronics and photonics,
Marinica {\em et al.} \cite{marinica}, Bulgakov and Sadreev
\cite{evgeny} and Prodanovi\'c {\em et al.} \cite{prodanovic}
reported the presence of BICs in photonic systems.
 Several mechanisms of formation of BICs in  open quantum dots (QDs) have been reported in the literature.
The simplest one is based on the symmetry of the systems and, as a consequence,
in the difference of parity between the QD eingenstates and the continuum spectrum \cite{texier}.
Another mechanism takes into account a nonzero coupling between bound states in the QD and the continuum spectra. The formation of BICs would be the result of a destructive interference process of these resonances, for certain variations of  the physical parameters of the QD \cite{Miyamoto, Sadreev1, Sadreev2}. A third mechanism for the BICs formation in optics, is associated with the Fabry-Perot interferometer\cite{Sadreev3}.

Until nowadays, there is only one experimental work, reported
by Capasso and \emph{co-workers} \cite{capasso}, in which BICs were measured
in semiconductor heterostructures grown by molecular beam epitaxy.
Thereby, the search of new systems which could be able to reveal the existence of BICs,
 with the possibility to be measured, is a very interesting and relevant field of research.
The experimental feasibility exhibits by graphene based systems and the  great advances in the controlled manipulation  and measurements reported in graphene, together with the possibility of modifying their electronic properties by applying external potentials, suggests that BICs could be observable in graphene quantum dots heterostructures.

\begin{figure}[ht]
\centering
\includegraphics[width=7.5cm] {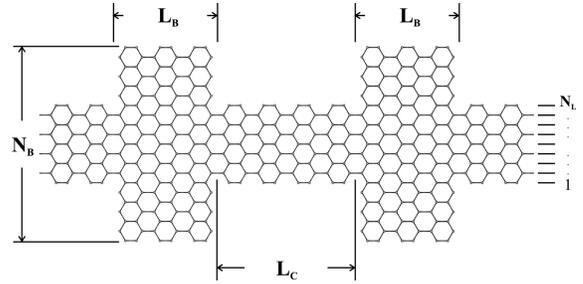}
\caption{Schematic view of a GQD structure with leads of width
$N_{L}=9$, a conductor region composed by two symmetrical
junctions of width $N_{B} = 21$ and length $L_{B} = 3$ separated by
a central structure of length $L_{C} = 4$ and width $N_{C} = 9$}
\label{fig1}
\end{figure}

In this work we study the formation of BICs in quantum-dot like
structures, formed by segments of graphene ribbons with different
widths connected between each other \cite{Gonzalez}. We identify the
presence of BICs in these symmetrical graphene quantum dots (GQDs)
and we discuss the mechanism for their formation. We found that the
GQD local density of states as a function of the energy shows
the presence of a variety of sharp peaks that we demonstrate are BICs.  The linear conductance also shows the
presence of  resonant states which contribute to the electronic
transmission.  By changing the geometrical parameters of the
structure, it is possible to control the number and position of
these resonances as a function of the Fermi energy.

A schematic view of the considered systems is presented
in figure \ref{fig1}. The conductor is formed by two symmetric
crossbar junctions of widths $N_{B}$ and length $L_{B}$, and a
central region that separates the junctions, of width $N_{c}$ and
length $L_{C}$.  Two semi-infinite leads of width $N_{L}=N_{c}$ are
connected to the ends of the central conductor. We studied the
different electronic states manifested in the system as a function
of the geometrical parameters of the GQD structure.

Systems are described by using a single $\pi$-band tight binding
Hamiltonian, taking into account first nearest neighbor interactions
with a hopping parameter $\gamma_0$. We consider hydrogen
passivation  by setting a different hopping parameter for the carbon
dimmers at the ribbons edges\cite{Son}, $\gamma _{edge} =
1.12\gamma_0$. To calculate  electronic properties of the system we
adopt the surface Green's functions matching formalism
\cite{Nardelli, Rosales}. In this scheme, we divide the
heterostructure in three parts, two leads composed by semi-infinite
pristine graphene nanoribbons, and the conductor region composed by two nanoribbon
crossbar junctions, as it is shown in figure \ref{fig1}.

In the linear response approach, the electronic conductance is
calculated by the Landauer formula. In terms of the conductor
Green's functions, it can be written as\cite{Datta} $ G =
\frac{{2e^2 }}{h}\bar T\left( {E } \right) = \frac{{2e^2 }}{h}
{\mathop{\rm Tr}\nolimits} \left[ {\Gamma _L G_C^R \Gamma _R G_C^A }
\right]$, where  $\bar T\left( {E } \right)$, is the transmission
function of an electron crossing the conductor region,
$\Gamma_{L/R}=i[ {\Sigma _{L/R} - \Sigma _{L/R} ^{\dag} }]$ is the
coupling between the conductor and the respective leads, given in
terms of the self-energy of each lead: $\Sigma _{L/R}
V_{C,L/R}\,g_{L/R}\,V_{L/R,C}$. Here, $V_{C,L/R}$ are the coupling
matrix elements and $g_{L/R}$ is the surface Green's function of the
corresponding lead \cite{Rosales}. The retarded (advanced) conductor
Green's functions are determined by \cite{Datta};$ G_{C}^{R,A}=[E -
H_C- \Sigma_{L}^{R,A}-\Sigma_{R}^{R,A}]^{-1}$, where $H_C$ is the
Hamiltonian of the conductor.

\begin{figure}[h]
\centering
\includegraphics [width=7.5cm] {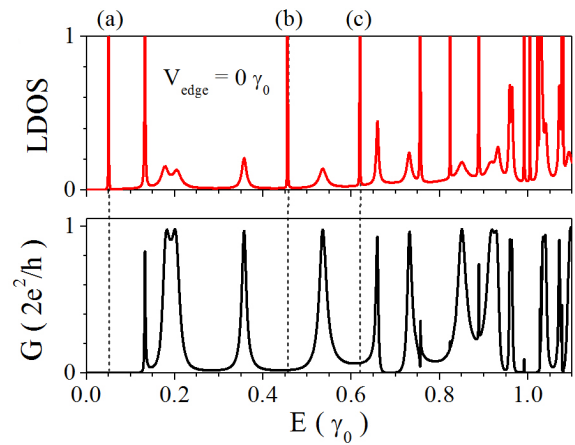}
\caption {LDOS (upper panel) and conductance (lower panel) as a
function of the Fermi energy for a GQD structure
based on leads of  width $N_L =5$, two
symmetric crossbar junctions of width $N_B = 17$ and  $L_B = 3$.
The central region has a width $N_C =5$ and length  $L_C = 5$. Marks (a), (b) and (c) denote position of peaks in the LDOS which are absent in the conductance. These states are identified as BICs.}\label{fig2}
\end{figure}

The Fig \ref{fig2} displays results of the local density os states (LDOS) (upper panel) and
the linear conductance (lower panel) for a GQD
structure formed by two armchair ribbons leads of width $N_{L}=5$
and a conductor region composed by two symmetric crossbar junctions
of width $N_{B}=17$, length $L_{B}=3$ and relative distances between
the junctions $L_C = 5$. The conductance of a pristine $N_L= 5$
armchair nanoribbon has been included for comparison (light green
dotted line).

It can be observed in the LDOS and in the conductance curves a series of peaks at
determined energies. This resonant
behavior of the electronic conductance arises from the interference
of the electronic wave functions inside the structure, which travel
forth and back forming stationary states in the conductor region
(well-like states).

\begin{figure}[h!]
\centering
\includegraphics[width=8.0cm] {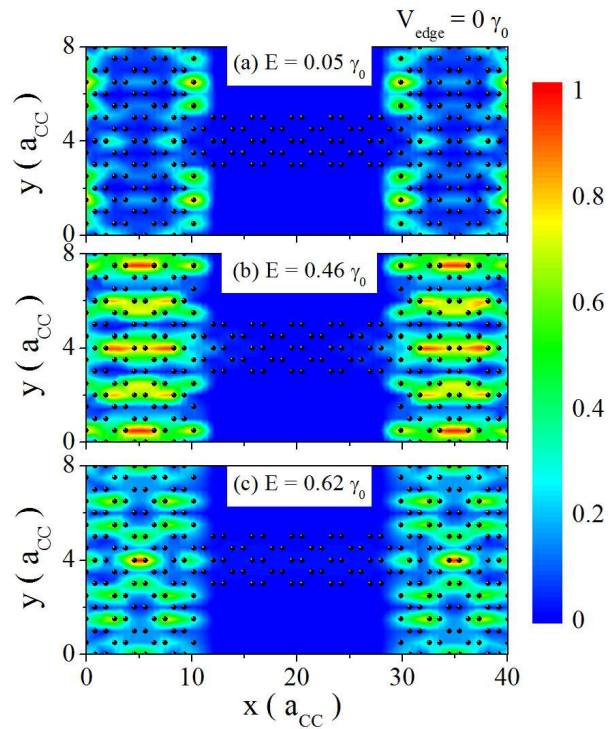}
\caption{Corresponding  contour plots of some sharp LDOS resonances marked in
  Fig. \ref{fig2} LDOS (upper panel).} \label{fig3}
\end{figure}

We start our analysis focusing on the sharp states present in the
LDOS displayed in this figure. We have marked the first three sharp
states with the letters (a), (b) and (c).  We note that the corresponding peaks in the conductance are absent and we
identify these states as BICs. We  calculated the spatial
distribution of these states, representing by the corresponding
contour plots exhibited in the figure \ref{fig3}. We observe that BICs are completely localized at the region defined by the crossbar junctions. Any electron from the leads,
injected at these energies, will be spatially
confined in the junctions due to the practically infinite (zero resonance width) lifetime of these states. Therefore BICs do not play any role in the transport properties of these GQD structures.
This can be seen in Fig. \ref{fig2}, where the corresponding
resonant peaks are absent in  the conductance curves.

\begin{figure}[h]
\centering
  \begin{picture}(220,100)(0,0)
    \thicklines
    \put(40,50){\line(1,0){140}}
    \put(110,50){\line(0,-1){30}}
    \put(110,50){\line(0,1){30}}
    \put(110,80){\circle*{12}}
   \put(110,20){\circle*{12}}
    \put(100,90){\makebox(0,0){$\varepsilon_{u}$}}
        \put(115,55){\makebox(0,0){0}}
    \put(100,10){\makebox(0,0){$\varepsilon_{d}$}}
    \put(120,30){\makebox(0,0){$V_{0}$}}
    \put(120,65){\makebox(0,0){$V_{0}$}}
    \put(170,55){\makebox(0,0)[b]{$\varepsilon_{w}=0$}}
    \put(40,50){\makebox(0,0)[r]{$.\, .\, .\, .\, .$}}
    \put(180,50){\makebox(0,0)[l]{$.\, .\, .\, .\, .$}}
    \multiput(50,50)(20,0){7}{\circle*{4}}
  \end{picture}
  \caption{Scheme of side-coupled impurities attached
            to a perfect wire.}
  \label{fig4}
\end{figure}
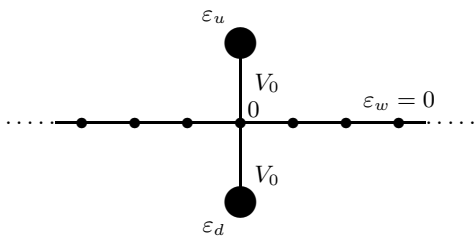

 We note that
the bound states in the continuum exhibit a spatial symmetry in the
transverse direction of the conductor, with the presence of nodes
and maximum in the amplitude of the  LDOS along that direction. This
behavior is a consequence of the crossbar junction symmetry with
respect to the longitudinal axis of the GQD structure. The localized
states belonging to the crossbar region could interact between each
other through the continuum of states of the pristine ribbon leading
to the formation of bound-states in the continuum.  In this sense, one of the
mechanisms of formation of BICS in our systems correspond to the first one
described in the introduction of this letter.
Follows the analysis of Moiseyev\cite{moiseyev}, the number of BICs can be
controlled by varying the gate potential applying in the up and down
barriers of the GQDs. In order to get a better understanding of this
phenomenon, we introduce a simple model that captures the essence of
the formation of BICs in our GQD structures. The model consists in two
side-coupled impurities attached to a perfect quantum wire
\cite{orellana} as shown in  Fig. \ref{fig4}.

By using the Dyson equation $G=g+gVG$ we calculate the Green's
function ($G$) in terms of the corresponding  Green's function of
the isolate sub-systems ($g$), here $V$ is the matrix coupling
between the impurities and the wire ($V_{0u}=V_{0d}=V_0$). To obtain
the LDOS of each impurity, $\rho_{\alpha}$ , ($\alpha=u,d$), we
calculate the imaginary part of the diagonal elements of the Green's
functions, $G_{\alpha}$. Setting the
site energies as, $\varepsilon_u=\varepsilon_0+\delta $ and
$\varepsilon_d=\varepsilon_0-\delta$, $\gamma=\pi V_{0}^2\rho(0)$, where $\rho(0)$ corresponds to the
LDOS in the site $0$ of the wire without impurities, and taking $\delta \ll \gamma$, the density of states of the entire system
is obtained summing over $\alpha$ and can be written approximately as,
\begin{eqnarray}
\rho&\approx&\frac{1}{\pi}\,\frac{2\gamma}
{(\omega-\varepsilon_0)^2+4\gamma^2}+\frac{1}{\pi}\,\frac{\delta
^2/2\gamma}{(\omega-\varepsilon_0)^2+(\delta^2/2\gamma)^2}.
\end{eqnarray}
The density of states is then the sum of two Lorentzian shapes lines
with widths $\Gamma_+ = 2\gamma$ and $\Gamma_{-}=\delta^2/2\gamma$,
corresponding to those states strongly and weakly coupled to the
continuum, respectively. In the limit of $\delta \rightarrow 0$,
$\Gamma_{-}$ vanishes and  the state weakly coupled to the
continuum becomes a bound-state in the continuum. This state arises
from the interference of the localized states in the impurities,
through the continuum states of the wire. In this sense, this
interference phenomenon is similar to phenomena like Fano and
Aharonov-Bohm effects. If we denote as $\psi_u$ and $\psi_d$ the
wave functions of the up and down impurity respectively, it is
straightforward to show that the antisymmetric
state,($\psi_u-\psi_d$) is an eigenstate of the complete system and
therefore it is a bound-state in the continuum.
\begin{figure}[ht]
\centering
\includegraphics[width=7.5cm] {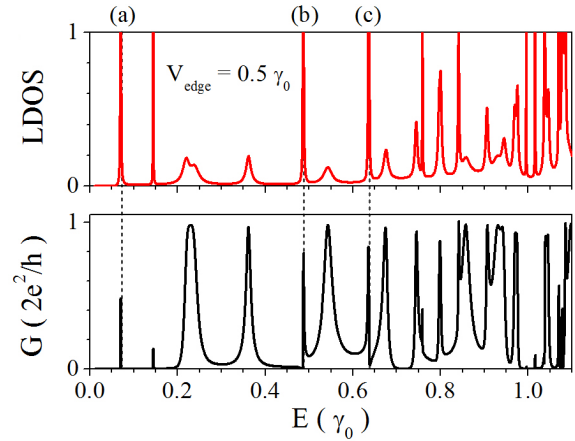}
\caption{LDOS (upper panel) and conductance (lower panel) as a
function of the Fermi energy for a GQD structure composed by the same parameters of Fig. \ref{fig2}, with an up-down
asymmetric gate potential applied to the junction regions.}
\label{fig5}
\end{figure}
In analogy, in the GQD
structure the formation of the BICs follows the same mechanism.
According to it, if any infinitesimal small perturbation breaks the
transversal symmetry, the BICs become resonant states with
infinitesimal widths. The widths of these "quasi-BICs" can be
controlled, for example, by tuning  the asymmetry of the system
through gate potentials. For instance, figure \ref{fig5} displays
the LDOS and conductance as a function of the Fermi energy, for the same
systems considered in figure \ref{fig2}, but now taking into account
a small up-down asymmetric gate potential applied to the edges of
the GQD. Due to this external perturbation, it is possible
to observe the apparition of new peaks of conductance at the BICs
energies levels in comparison with the unperturbed case. Therefore,
for the feasibility to observe this phenomenon, it is necessary to
build a GQD as symmetrical as possible and to control the asymmetry
via gate potentials. By measuring the changes in the widths of the
quasi-BICs in the conductance, it would be possible to obtain
indirect evidence of the BICs.

In summary, in this work we have studied the formation of
bound-sates in the continuum in quantum-dot like structures. We
identify the presence of these states in the LDOS in symmetrical
graphene quantum dots structures and we discuss the mechanism of
formation of these kind of exotic states. Our results suggests that BICs could be observable in GQDs.
The BICs or quasi BICs can
have applications in the fields of the spintronics and the quantum computing. For
instance, by applying a magnetic field to the GQD structure, the
spin degeneracy is lifted and by controlling the resonances width
with a gate potential, an efficient spin filter can be designed.
On
the other hand, the presence of two simultaneous BICs could be used
as a qubits for quantum information. Also,
BICs could have important applications in photonic crystals, so in this
sense, we expect that the study of this kind of states in graphene
nanoribbons quantum dots could open a new line of applied
research.

The authors acknowledge the financial support of CONICYT/Programa
Bicentenario de Ciencia y Tecnolog\'{\i}a (CENAVA, grant ACT27),
USM 110971 internal grant, FONDECYT program
grants 11090212, 1100560 and 1100672. L. R. also acknowledges to PUCV-DII grant 123.707/2010.

\end{document}